# Percolation in photonic crystals revealed by Fano Resonance

Jose Angel Pariente,[1] Farzaneh Bayat,[1,2] Carlos Pecharomán,[1] Alvaro Blanco,[1] Antonio García-Martín,[3] Cefe López[1]

[1] Instituto de Ciencia de Materiales de Madrid (ICMM); Consejo Superior de Investigaciones Científicas (CSIC) Calle Sor Juana Inés de la Cruz 3, E-28049 Madrid Spain, c.lopez@csic.es
[2] Photonics Group, Research Institute for Applied Physics and Astronomy, University of Tabriz, Tabriz, Iran
[3] Instituto de Microelectrónica de Madrid (IMM), Consejo Superior de Investigaciones Científicas (CSIC) Isaac Newton 8 (PTM), E-28760 Tres Cantos, Madrid, Spain

The understanding of how the arrangement of defects in photonic crystals impacts its photonic properties is crucial for the design of functional materials based thereon. By preparing photonic crystals with random missing scatterers we create crystals where disorder is embodied as vacancies in an otherwise perfect lattice rather than the usual positional or size disorder. We show that the amount of defects not only determines the intensity but also the nature of the light scattering. As the amount of defects varies, light scattering undergoes a transition whereby the usual signatures of photonic gaps (Bragg peak) suffer line-shape changes (Bragg dip) that can be readily described with the Fano resonance $q$ parameter. When the amount of vacancies reaches the percolation threshold, $q$ undergoes a sign change signaling the transition from a crystal to a mosaic of microcrystals through a state where scattering is maximum. Beyond that point the system reenters a state of low scattering that appears in the guise of normal Bragg diffraction.

The control of light and its interaction with matter has a huge impact on technology and has taken materials design well beyond subwavelength range. Two of the most paradigmatic examples of structured material platforms to study light-matter interaction are photonic crystals[1,2] (PhC) and photonic glasses[3] (PhG). The difference between them lies in the nature of the arrangement of scattering elements which is in the origin of the different forms in which scattering is incarnated: PhCs building up photonic bands (which requires translational symmetry and entails order) and PhGs through diffusion (which is the general case when disorder[4] is present). There is a subtractive way to obtain a PhG from a crystal: creating random vacancies in a PhC by randomly removing scattering elements.[5] Depending on the quantity of vacancies, the final structure may appear bereft of any type of periodicity.

It is natural to expect that the optical properties of a PhC deteriorate by the inclusion of disorder and suffer some transformation from showing features associated with photonic bands to showing those from diffusion.[6] However, if the process is carried out by removing scattering elements without altering the underlying lattice, it comes as no surprise that for large concentrations of vacancies the remaining elements become a *minority* in an otherwise homogeneous medium. The latter process would lead the system away from diffusive again.

During this transition, there is necessarily a critical concentration for which one cluster of vacancies connects opposite edges of the structure while, at the same time, these edges are also connected by the remaining scatterers network. This is the definition of percolation which can be found in many networks or particulate systems and has most impact regarding conduction phenomena,[7,8] although several examples can also be found in other fields such as: medicine,[9] economics,[10] sociology[11] or geography[12] to name a few. In this work, the site percolation for an *fcc* lattice PhCs[13] (artificial opal) is studied in relation with optical scattering. Great effort[14,15,16] has been devoted to accurately calculate the site percolation threshold for an *fcc* lattice and many methods cast values that, on average, are around: $\varrho_c$ = 20%. While the onset of percolation has very intuitive meaning in terms of electrical conductivity and related phenomena such as the optical response of metallic layers that is enormously enhanced by the fractal character of the sub monolayer coverage[17,18] little has been studied in connection with the optical properties of dielectric-only composites.[19]

Fano resonance takes place in systems where an interaction couples a discrete state and a continuum of states. Ugo Fano was first to observe this resonance and their asymmetric profiles in the auto-ionization cross section of noble gases[20] and later developed a quantum theory[21] that has been found pervasive across many disciplines in physical sciences. From that day on, Fano resonance has been studied in the multitude of physical systems.[22]

In general light transport processes, when the input and output ports are coupled both directly (continuum of states) and through a resonant cavity (discrete state), the resulting response presents a Fano resonance. If the direct coupling is not perfect, *i.e.*, partially reflecting even without the cavity mode, then one obtains an asymmetric spectrum with a sharp peak followed or preceded by a sharp dip.[23] The asymmetry of the peak indicates the relative contribution of either channel and enters the cross section expression through the parameter $q$. Fano resonance has been observed in slabs,[24] spiral[25] and disordered[26] PhCs and in opals changing the dielectric constant was used to set the coupling between the band gap (discrete) and the Mie (continuous) scattering background.[27]

In studying the optical response $fcc$ photonic crystals as a function of the fraction of missing spheres (under the constraint that those remaining lie on lattice sites) we have found that specular reflectance presents a Fano-like line-shape whose principal parameter, $q$, can be tuned through the vacancy fraction and can be made to change sign at crossing the percolation threshold.

Changing the fraction of vacancies changes the balance between the two possible scattering channels (band gap and background). So, it is possible to choose the type of asymmetric spectrum and refer it to percolation. At the percolation threshold, as a result of the divergence of cluster size, $\xi(p) \sim |p - p_c|^{-\nu}$ and cluster boundary (enhanced by fractal dimension), scattering is essentially determined by the surface and the resonant character of Bragg diffraction is all but lost. The effect of the band structure is inhibited giving rise to an enhancement of non-resonant scattering.

To do so, binary mixtures of monodisperse Polymethyl Methacrylate (PMMA) and Polystyrene (PS) spheres (acting here as dopants) of 334 and 313 nm of diameter respectively were used to prepare different 3D PhCs with a controlled amount of vacancies by the standard vertical deposition method.[28] Once the final structures were achieved (**Figure** 1a), selective etching of the PS dopants yielded opals with a precisely controlled amount of vacancies from 0% to 50%. In **Figure** 1b we can appreciate that, after the etching, the PMMA spheres are not affected by the cyclohexane. An $fcc$ structure containing a random distribution of the vacancies can be seen. A fine control of the vacancies in the opal is achieved by tuning the PS concentration in the colloidal suspension. Figure 1c shows the calibration of the vacancies content obtained by counting missing spheres from SEM images and relating them to colloids used to prepare them. The actual content is consistently 25% larger than the amount of PS in the colloidal preparations; this small deviation is attributed to the error in the nominal concentration of the colloids and has been taken into account to correct the vacancy fraction. Moreover, SEM inspection of internal facets obtained by cleaving the samples, reveals that the distribution of vacancies inside the structure is uniform (Figure 1d).

Fortunately, the percolation of vacancies is expected at low enough filling fractions so as not to compromise the mechanical stability of the crystal. Thus, we have been able to produce samples with filling fractions well beyond the site percolation threshold. At this point we must mention that the system is subjected to a mechanical restriction in that all particles in the final structure must remain connected. Otherwise, isolated (floating) particles or clusters could result when all the particles around them were etched. In this event they would be pulled into contact with other particles by gravity or electrostatic forces.

Optical characterization was carried out by measuring both specular reflectance and transmittance in a standard Fourier Transform Infrared spectrometer at normal incidence (perpendicular to the opal (111) planes). For the sake of clarity, only reflectance will be discussed, keeping in mind that the same results are obtained for the absorbance.

**Figure 2** shows a contour plot containing the reflectance spectra as a function of wavelength and vacancy fraction in the range 0-38% for 20 layers thick structures, a safe choice of thickness to be able to consider the samples infinite. One of the most remarkable features of the disorder present in this kind of system is that, at variance disorder induced by polydispersity of the spheres, long range correlation is here preserved. One is tempted to think that even when the fraction of missing scatterers is high, scattering by microcrystals is coherent.[29] In fact a rough estimate of Bragg wavelength based on average refractive index that can be approximated as $\langle n^2 \rangle = 0.26 + 0.74[n_{PMMA}^2 - \rho(n_{PMMA}^2 - 1)]$ where ϱ is the fraction of missing spheres predicts a decreasing behavior of the Bragg peak that is not observed in the experiments for concentration as high as 10% or more. Here, 3 very different regions can be clearly identified at first sight: region 1, from 0 to 19 % vacancies and below percolation threshold, region 2, from 20 to 25% vacancies, within percolation threshold, and region 3, above 25% vacancies and well beyond percolation threshold. Thinner samples yielded different behavior due to finite size effects whose analysis will be provided elsewhere. Reflectance spectra belonging to any of these three regions can be fitted to different Fano line-shapes according to the following general formula:

$$F(\epsilon) = A + B\frac{(\epsilon+q)^2}{1+\epsilon^2} \qquad (1)$$

where a single parameter, $q$, governs the asymmetry. $A$ and $B$ are constants related to background and intensity, $\epsilon = (\omega - \omega_0)/(\gamma/2)$ is the dimensionless frequency related to the position, $\omega_0$ and $\gamma$ are the central frequency and the width of the resonance.

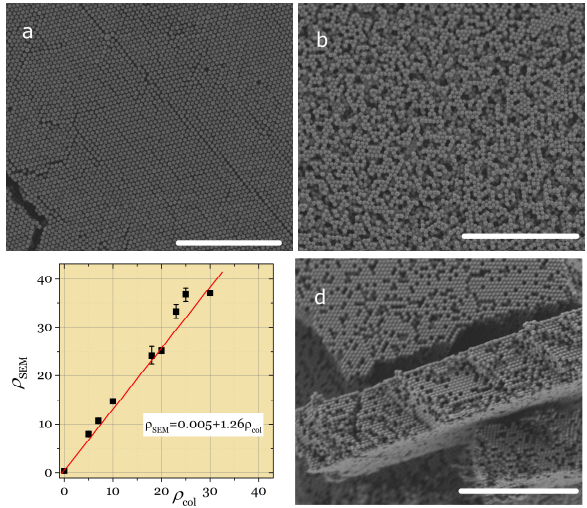

Figure 1. SEM images for different types of vacancy-doped photonic crystals grown by vertical deposition. (a) PMMA and PS spheres of 334nm are used as blocks for the opals. (b) Removing the PS with cyclohexane, a random distribution of vacancies is performed in the whole opal. These two images are for a 50% of PS/vacancy doping. (c) Statistics of the vacancies for some samples versus the theoretical vacancies. An error of 25% is committed introducing vacancies. (d) Cross section for VPhC of ρ = 25%, the vacancies are homogeneously distributed in the whole sample. The scale bar in the SEM images is 10μm.

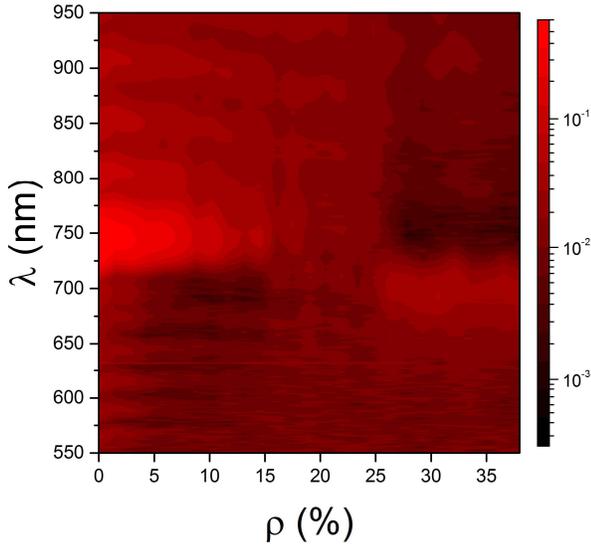

Figure 2. Surface contour plot of reflectance spectra for 20 layers increasing the number of vacancies. On the percolation threshold ($\rho_c$ = 20-25%), the band gap disappears due to the effect of the infinite scattering provided by the cluster of vacancies. At $\rho < \rho_c$, the scattering of the vacancies reduces the intensity of the Bragg peak and by the coupling with Bragg scattering an asymmetric spectra are obtained. At $\rho > \rho_c$, the band gap is shifted due to the contribution of the vacancies but the band structure does not disappear.

**Figure** 3 shows reflectance spectra from a 20 layers thick PhC at four representative vacancy concentrations and their respective fits to a Fano line-shape, in order to obtain the $q$ parameter. In this work, three regions are distinguished. Well below the percolation threshold ($\rho \ll \rho_c$) a lorentzian profile (**Figure** 3, $\rho$ = 0%) is obtained. Since the gap width is rather large the fitting for very low $\rho$ is not perfect. This line-shape turns into an asymmetric profile (**Figure** 3, $\rho$ = 13%) as the threshold concentration is approached. The central wavelength of the band gap does not shift during this process. Around or at the percolation threshold ($\rho = \rho_c$) a weak dip at the wavelength of the band gap can be seen. In terms of optical properties, the net effect is that a band gap due to the band structure developed into a pass band where reflectance is inhibited (**Figure** 3, $\rho$ = 26%). Above the percolation threshold ($\rho \gg \rho_c$) an asymmetric profile is recovered that should tend to a lorentzian profile far above the threshold. Notice that the peak is the mirror image of that below threshold with the relative maxima and minima positions exchanged (**Figure** 3, $\rho$ = 32%). This line-shape change can be obtained simply through a change of sign in the $q$ parameter of the Fano resonance as discussed below. The proposed singular system, in which spheres are removed without altering the lattice parameter, preserves band gap features even for a high percentage of the vacancies.

Fano resonance is expected when a coupling between a discrete state (narrow band) and a continuum states (broadband) exits. In this study, the band gap acts as the narrow band due to Bragg diffraction. This range of wave vectors and prohibited frequencies are sufficiently narrow in comparison with the whole frequencies available for the photons. The broadband can be identified as the Mie scattering background brought about as a result of the presence of the vacancies. The interference between these two scattering paths gives rise to the Fano resonance (**Figure** 3).

In this context, in region 1 the intensity of the Bragg peak (centered round 740 nm) is reduced as the number of vacancies increase. Samples belonging to this region present a behavior typical of photonic crystals with defects. A lorentzian peak is measured corresponding to Fano line-shapes with $q$ parameter tending to infinity. In this case, there is little or no contribution of scattering through the continuum. Therefore, the only possible transition is through the (modified) discrete state. This is consistent with previous works on diffusion[5,30] and optical loss in photonic crystals.[31,32] From 10 to 19%, the intensity of the Bragg peak keeps decreasing indicating that the Mie scattering grows stronger at the expense of Bragg scattering as a result of increased vacancies fraction. The scattering through background brings the $q$ parameter close to unity so the interference between Bragg and Mie scattering gives rise to very asymmetric spectra (low $\rho$ region 1 in Figure 2, and fit shown in **Figure** 3, $\rho$ = 13%).

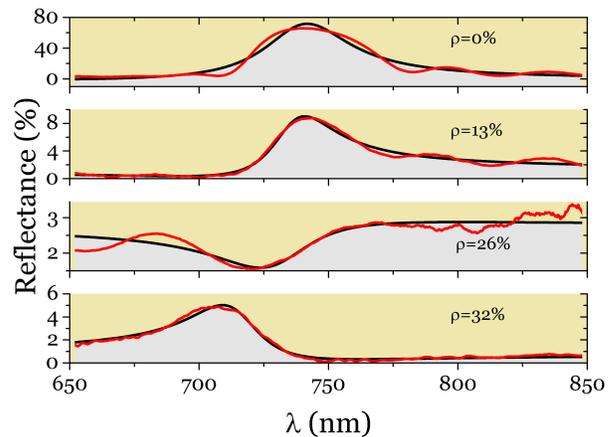

Figure 3. Reflectance spectra (red line) and the fit (black line) to the Fano equation at the band gap for (a) 0%, (b) 10%; (c) 20% and (d) 25%. Different symmetric or asymmetric profiles are obtained depending on the percentage of vacancies. The number of the layers in all the spectra are 20.

From 20 to 25%, the Bragg peak turns into a weak dip (intermediate ϱ in Figure 2, and fit shown in **Figure 3**, ϱ = 26%). The band gap features disappear and, as a result of thin film interference, only Fabry-Perot fringes can be seen. This is the fraction where the percolation threshold is expected for an *fcc* structure. It means that, in all likelihood, at least one cluster of vacancies spans the sample from edge to edge. By means of percolation, a huge enhancement of the Mie scattering is expected in the sample because this singular concentration is such that both larger or smaller concentrations simply drive the system towards a more homogeneous one. Obviously lower concentrations (ϱ → 0) tend to the perfect photonic crystal while larger concentrations (ϱ → 1) simply represent an emptier and therefore also more uniform system where vacuum is sparsely decorated with the remaining scattering particles. In addition, it should be noticed that near percolation most of the magnitudes that are related to the clusters, such as their size or the size of their boundaries, diverge according to power laws: $s \sim |\rho - \rho_c|^{-\nu}$ with $\nu > 0$ the critical exponent.[33] Thus, an infinite scattering is provided allowing all frequencies to propagate. The immediate consequence of this fact is not only the inhibition of the band structure of the sample, but also the improvement of the transport through it.

In the range of concentrations above the threshold the divergence of the diffuse scattering due to percolation lessens and asymmetric spectra are recovered only with an inversion in the asymmetry: the dip is now to the right of the maximum (high ϱ region in Figure 2, and fit shown in **Figure 3**, ϱ = 32%). This is indicative of a change of the sign in the *q* parameter. In this case, for the spectra above the percolation threshold a negative *q* value is obtained.

The value of *q* describes the relative importance of the intervening channels and derives from the strength of the interaction with the perturbation introduced by the continuum in the perturbative approximation of the original quantum formulation. In our case, it is determined by the concentration of the vacancies in the opals and a value of zero points to a total dominance of non-resonant scattering as opposed to photonic band-mediated where *q* would tend to infinity.

The behavior described previously can be seen in **Figure 4** that summarizes the results concerning the *q* parameter obtained from the fittings to the Fano equation (1) for different VPhC for all concentrations and all thicknesses studied that range from one to 35 monolayers. This type of representation is most appropriate since it consolidates data spread owing to normal statistical dispersion and that due to thickness dependence. The whiskers signal the maximum and minimum values of the *q* parameter for each percentage; the white lines are the mean values and the green boxes mark one standard deviation from the mean values. If we assume that the *q* parameter is close to zero and changes sign in percolation and it has a positive (negative) value below (above) percolation threshold we can state that percolation takes place at $\varrho_c \approx 22 \pm 3\%$. The coupling between Bragg and Mie scattering is stronger when we add vacancies to the photonic crystal. In the percolation neighbourhood (pink shaded area in **Figure 4** marks the range -1< *q* <1), scattering by clusters of vacancies is the dominant process so the scattering through band structure of the photonic crystal becomes negligible. Beyond this region, the overwhelming contribution of background scattering is diminished because cluster boundary size decreases causing a recovery of the band structure scattering. A negative *q* value is obtained in this region giving rise to a blue shifted peak with respect to the band gap of the initial crystal.

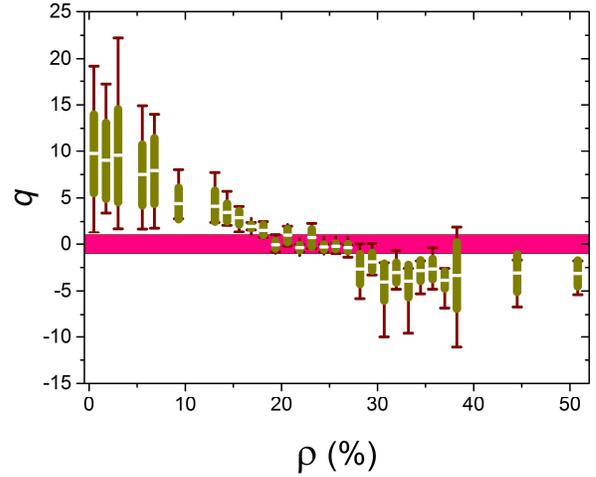

Figure 4. *q* parameter as a function of the vacancies for different number of layers comprising all the thicknesses measured: from one to over thirty monolayers. The percentage of vacancies determines the intensity of the interaction between the band gap and the Mie scattering background. According to this percentage, positive ($\rho < \rho_c$), negative ($\rho > \rho_c$) or close to zero ($\rho = \rho_c$, pink shaded area) *q* values are obtained. The whiskers represent the maximum and minimum obtained *q* parameter. The boxes are the standard deviation and the white lines are the mean from the values.

In conclusion, in this work we have studied, by a fine control over extrinsic vacancies added, the variation of optical properties for different percolation regimes in photonic crystals with vacancies. Introducing vacancies, the balance between the relative intensity of scattering through band gap and through Mie scattering background takes place giving rise to a Fano resonance manifest in the optical spectra. Not only have we achieved all the different asymmetric profile characteristics of the Fano resonance, but we have also set a correlation between the *q* parameter and the percentage of vacancies. At the geometric percolation concentration of the vacancies in an opal, the *q* parameter is close to zero. This agrees with the diverging size of the percolated cluster which gives rise to an intense Mie scattering background inhibiting the band gap and annihilating ballistic reflectance and transmittance at the same time.